\newif\ifuseprd
\newif\ifom
\useprdfalse 
\omfalse    
\ifuseprd 
\documentclass[preprint,tightenlines,floats,prd,eqsecnum,nobibnotes,%
nofootinbib]{revtex4}
\else
\documentclass[final,12pt,letterpaper]{JHEP}
\fi 
\usepackage{amsmath,amssymb}
\usepackage[final]{epsfig}

\newif\ifspinpm 
\spinpmtrue 

\allowdisplaybreaks[3]
\ifom
\input dnmacs
\fi 

\def\omt{{\ifom{{\dn\dnhalf :}}\else%
        {{3\!{\footnotesize$\mathbf{{\frown}\llap{\text{\tiny$\prime$}}}$}%
        {\hbox to -.7ex{\null}\llap{\raise1.3ex\hbox{\tiny{%
        \setbox255=\hbox{$\mathbf{{\smile}}$}%
        \copy255\kern-.7\wd255{\raise.5ex\hbox{$\mathbf{\cdot}$}}}}}}}}\fi}}

\newcommand\skipthis[1]{{}}

\let\oldAE\AE
\renewcommand\AE{{\ifmmode{\text{\it\oldAE}}\else{\oldAE}\fi}}

\newcommand\ct[1]{{\ifuseprd{\em{#1}},\else{\sf {#1}},\fi}}

\chardef\til=`~

\newcommand\ep{\epsilon}

\providecommand\putabstract[1]{\ifuseprd\begin{abstract} {#1} \end{abstract}%
                           \else \abstract{{#1}} \fi}
\providecommand\plb[3]{{Phys.\ Lett.\ B {\bf {#1}}, {#3} ({#2})}}
\providecommand\npb[3]{{Nucl.\ Phys.\ {\bf B{#1}}, {#3} ({#2})}}
\providecommand\jhep[3]{{J.\ High Energy Phys.\ {\bf #1}, {#3} ({#2})}}

\ifuseprd
\begin{document} 
\fi 

\title{Einstein-Hilbert action on the brane for the bulk graviton} 
\ifuseprd
\author{Steven Corley}\email{scorley@het.brown.edu}
\author{David A. Lowe}\email{lowe@het.brown.edu}
\affiliation{Department of Physics \\ 
       Brown University \\
       Providence, RI \ 02912 \\
       USA}
\else 
\author{Steven Corley,\thanks{\tt scorley@het.brown.edu} \
David A. Lowe \thanks{\tt lowe@het.brown.edu} \
and Sanjaye Ramgoolam\thanks{\tt ramgosk@het.brown.edu} \\
Department of Physics\\
Brown University \\
Providence, RI \ 02912\\
USA}
\fi 

\putabstract{
We find new closed string couplings on Dp-branes for the
bosonic string.  These couplings are quadratic in derivatives
and therefore take the form of induced kinetic terms on 
the brane.  For the graviton in particular we find the
induced Einstein-Hilbert term as well as terms quadratic
in the second fundamental tensor. We comment on tachyon dependences 
of these brane-localized couplings. 
}

\preprint{BROWN-HET-1267\ifuseprd,~\else\\\fi }

\ifuseprd
\maketitle
\else
\begin{document}
\fi 

\section{Introduction}

$D$-branes have played a prominent role in string theory
since the realization that they are the carriers of
Ramond-Ramond (or RR) charge arising in the superstring.  
They are described in the sigma-model context by submanifolds
where open strings end.  From the spacetime point of view, and
at low energies compared to the string scale, 
they are submanifolds on which massless fields 
propagate and interact
with the bulk, or closed string, fields.  To leading
order in an $\alpha^{\prime}$ expansion with respect
to the closed string fields (to zeroth order in derivatives on the
closed string fields) the low energy effective
action governing the $D$-brane - bulk interactions
for a single brane
is known to be described by the Born-Infeld action
plus Wess-Zumino terms.

A natural question to ask is what are the next order
corrections to the Born-Infeld low energy effective
action.  These corrections are quadratic
in derivatives acting on the closed string fields.
Such terms would consist of kinetic-like terms for
the pull-backs of the closed string fields to the
brane.  For example there could be an Einstein-Hilbert
term for the pullback of the metric or the square
of the field strength for the pullback of the
NS-NS two-form field.  Actually though there are
many other possible terms as well, eg., the square
of the second fundamental term (a generalization of
the extrinsic curvature for codimension $>1$ branes).
It is known, for example, that a single BPS $D$-brane
in Type II string theory does not have such terms for the metric \cite{BBG}.
This is easy to see from the string scattering
amplitudes evaluated in \cite{KT,GarMye1} for scattering
a graviton off a $D$-brane.  
The point
is simply that these
amplitudes have no quadratic dependence on the momenta.
The field theory terms that we are looking for however
contribute to the graviton-graviton amplitude at 
quadratic order, and they are the only terms to
contribute at this order, so one concludes that they
cannot be in the low energy effective action (there
are however quartic in derivative, or $(\alpha')^2$,
corrections, see eg. \cite{BBG,Wyl,Fot}) .

Our purpose here is to show that this argument no 
longer holds for the $D$-branes appearing in the
bosonic string\footnote{For the space filling $D25$
branes see \cite{aat1,aat2}}.  Specifically we show that kinetic-like
terms for the massless closed string fields are 
required in the $D$-brane low energy effective action
in order to reproduce the string scattering amplitudes.
This includes in particular the Einstein-Hilbert
term for the induced metric on the brane, as well
as kinetic terms for the pullbacks of the antisymmetric
tensor and dilaton.  

Interest in the induced Einstein-Hilbert term
has appeared recently in the work
of \cite{Dvali}.  Specifically it was argued in
\cite{Dvali} that such a term would give rise
to a gravitational interaction between matter
on the brane that goes like $1/r^{p-2}$ for
a $p$-dimensional brane as opposed to the
naive expectation of $1/r^{D-3}$ for a $D$ dimensional
spacetime due to the bulk gravitational force.
This argument presupposes that the brane is
infinitely thin.  It was further argued however \cite{Dvali2}
that a brane with finite thickness would also
exhibit the $1/r^{p-2}$ gravitational potential on
the brane up to some crossover point, and that
only for larger separations would the potential
become $1/r^{D-3}$.  

The outline of the paper is as follows.  In section 2
we discuss the tree level string amplitude computation
corresponding to scattering a massless closed string
field off a bosonic $Dp$-brane.  In section 3 we find
the terms needed (in addition to the Born-Infeld action)
in the low energy effective action to reproduce 
these string amplitudes.  In particular we find that 
terms quadratic in derivatives acting on the closed string
fields are necessary.  In section 4 we discuss ambiguities
in our brane terms arising from the freedom in making
field redefinitions.  In section 5 we discuss the corrections these
terms make to the graviton propagator, and connections with the work
of \cite{Dvali}. In section 6, we comment on couplings to the
open-string tachyon following \cite{sen,sen2}.

\section{String scattering amplitudes}

In this section we compute the tree level bosonic string scattering
amplitude associated with scattering a massless closed string excitation
off a $Dp$-brane.  The tree level computation involves computing
correlators of pairs of closed string fields on the disk.  

Recall that a $Dp$-brane is defined
in the sigma-model context as a $(p+1)$-dimensional submanifold
where open strings end, eg., for a $p$-brane localized in
the $x^{p+1},...,x^{D}$ directions the open string endpoints satisfy
\begin{eqnarray}
\partial_{\tau} X^{a} |_{\sigma  =  0,\pi} & = & 0 \, ; \,\, a =
0,...,p~, \nonumber\\
X^{i} |_{\sigma = 0, \pi} & = & x^{i} \, ; \,\, i=(p+1),...,D~,
\end{eqnarray}
where $D=26$ for the bosonic string.  The two-point correlator of
string coordinates $X^{\mu}(z,\bar{z})$ for such boundary conditions
on the disk is given by
\begin{equation}
\langle X^{\mu}(z,\bar{z}) X^{\nu}(w, \bar{w}) \rangle =
- \frac{\alpha'}{2} \left(\eta^{\mu \nu} \ln |z-w|^2
- V^{\mu \nu} \ln |z - \bar{w}|^2 \right)~,
\label{2point}
\end{equation}
where $\eta^{\mu \nu}= N^{\mu \nu} + D^{\mu \nu}$
and $V^{\mu \nu} = D^{\mu \nu} - N^{\mu \nu}$
with $N^{\mu \nu} = \mbox{diag}(-1,1,...,1,0,...,0)$ where the last
1 occurs at the $p^{th}$ diagonal index and 
$D^{\mu \nu} = \mbox{diag}(0,...,0,1,...,1)$ where the first 1 occurs
at the $(p+1)^{th}$ diagonal index.  The notation is chosen so 
that $N^{\mu \nu}$ corresponds to the metric in the Neumann
boundary conditions directions and $D^{\mu \nu}$ to the metric
in the Dirichlet boundary conditions directions.  
Note that $\eta^{\mu \nu}$ is just the Minkowski metric.

The fields that we shall be interested in are the graviton,
antisymmetric tensor, and dilaton.
The vertex operators for these fields are all described
by (splitting the string
coordinates into holomorphic and antiholomorphic pieces
as $X^{\mu}(z,\bar{z}) = X^{\mu}(z) + \tilde{X}^{\mu}(\bar{z})$)
\begin{equation}
V_{\epsilon} = \frac{2}{\alpha'} \epsilon_{\mu \nu}
 :\partial X^{\mu}(z) e^{i k \cdot X(z)}: 
:\bar{\partial} \tilde{X}^{\nu}(\bar{z}) e^{i k \cdot
 \tilde{X}(\bar{z})}:~,
\end{equation}
where the colons denote separate normal ordering for
the left and right moving fields.
On shell the momentum and polarization tensor satisfy
the  massless and transverse conditions, $k^2 = 0 =
k^{\mu} \epsilon_{\mu \nu}$, as well as the 
following conditions for the respective vertex operators:
\begin{eqnarray}
\epsilon_{\mu \nu} & = & \epsilon_{\nu \mu},\,\, 
\epsilon^{\mu}_{\,\,\,\,\mu} =0, \,\,\,\, \mbox{graviton}, \nonumber\\
\epsilon_{\mu \nu} & = - & \epsilon_{\nu \mu},\,\,\,\,
  \mbox{antisymmetric tensor}, \nonumber\\
\epsilon_{\mu \nu} & = & \frac{1}{\sqrt{D-2}} (\eta_{\mu \nu}
- k_{\mu} l_{\nu} - l_{\mu} k_{\nu}), \,\, k^{\mu} l_{\mu} = 1, 
\,\,\,\,\,\, \mbox{dilaton}.
\end{eqnarray}

The amplitude associated with scattering any one of these
closed string fields off a $Dp$-brane (or converting one of
these fields into a possibly different field) is 
given by
\begin{equation}
{\cal A}_{D_2}(k_1, \epsilon_1; k_2, \epsilon_2) = g^{2}_c 
e^{-\lambda} \int^{1}_{0} dy \langle c(z) V_{\epsilon_1}(z,\bar{z})
c(w) \tilde{c}(\bar{w}) V_{\epsilon_2}(w,\bar{w}) \rangle 
|_{z=iy, w=i}~,
\label{scatamp}
\end{equation}
where $g_{c}$ is the closed string coupling constant, $\lambda$ the
dilaton vev, and the $c$'s are the ghost fields arising from 
fixing the $SL(2,R)$ symmetry of the disk.  The ghost correlator,
which provides the necessary gauge fixing measure,
is straightforward to evaluate on the disk and is given by
\begin{equation}
\langle c(z) c(w) \tilde{c}(\bar{w}) \rangle = C^{g}_{D_2} (z-w)
(z - \bar{w})
(w-\bar{w})~,
\label{ghostcorr}
\end{equation}
where $C^{g}_{D_2}$ is a functional determinant whose value (times
some other factors and functional determinants) will
be fixed by comparing to the Born-Infeld low energy effective
action.

The remaining piece of the integrand (\ref{scatamp}) to be evaluated
is the correlator of massless closed string vertex operators.
The computation is straightforward but tedious and we find
\begin{eqnarray}
\langle V_{\epsilon_1}(z,\bar{z}) V_{\epsilon_2}(w,\bar{w}) \rangle
& = & i \, (2 \pi)^{p+1} \, \delta^{p+1}(k_1 + k_2) \, C^{X}_{D_2} \,
|z-w|^{2k_1 \, k_2} |z-\bar{w}|^{-2 k_1 V k_2} \nonumber \\
& \cdot & |z-\bar{z}|^{-k_1 V k_1} \,
|w-\bar{w}|^{-k_2 V k_2} \biggl[
\frac{d_1}{|z-w|^2 |z-\bar{w}|^2} \nonumber \\ & - & \frac{1}{(z-\bar{z})
  (w-\bar{w})} \Bigl( 
\frac{d_2}{|z-w|^2} -
\frac{d_3}{|z-\bar{w}|^2} \Bigr)
+ \frac{d_4}{|z-\bar{w}|^4} \nonumber \\ & + & \frac{d_5}{|z-w|^4} 
+ \frac{1}{|z-\bar{z}|^2 |w-\bar{w}|^2} \Bigl(d_6+ d_7 \frac{|z-\bar{w}|^4}
{|z-w|^4} \nonumber \\
& - & d_8 \frac{|z-\bar{w}|^2}{|z-w|^2} - d_9
\frac{|z-w|^2}{|z-\bar{w}|^2} + d_{10} \frac{|z-w|^4}{|z-\bar{w}|^4}
\Bigr) \biggr]~,
\label{masslesscorr}
\end{eqnarray}
where $C^{X}_{D_2}$ is a functional determinant and
the coefficients are given by
\begin{eqnarray}
d_1 & = & \biggl[ \frac{1}{2} k_2 \ep^{T}_1 \ep_2 k_1 
+ \frac{1}{2} k_2 \ep_1 \ep^{T}_2 k_1 
 - k_2 \ep_1 V \ep_2 k_1 - k_2 \ep^{T}_1
\ep_2 V k_2 - k_2 \ep_1 \ep^{T}_2 V k_2 \nonumber \\
& + & (\mbox{1} \leftrightarrow
\mbox{2}) \biggr] \nonumber\\
d_2 & = & \mbox{Tr}(\ep_1 V) (k_1 \ep_2 k_1) +(\mbox{1} \leftrightarrow
\mbox{2}) \nonumber\\
d_3 & = & k_1 V \ep_1 V \ep_2 V k_2  + 
\frac{1}{2} k_1 V \ep^{T}_1 \ep_2 V k_2  + 
\frac{1}{2} k_1 V \ep_1 \ep^{T}_2 V k_2 \nonumber \\
& + & \mbox{Tr}(\ep_1 V)
(k_1 V \ep_2 V k_1)  +(\mbox{1} \leftrightarrow
\mbox{2}) \nonumber\\
d_4 & = & \mbox{Tr}(\ep_1 V \ep_2 V) + k_1 V \ep_2 V \ep_1 V k_2
+ k_2 V \ep_1 V \ep_2 V k_1 \nonumber\\
d_5 & = & \mbox{Tr}(\ep_1 \ep^{T}_{2}) - k_1 \ep^{T}_2 \ep_1 k_2
- k_2 \ep_1 \ep^{T}_2 k_1 \nonumber \\
d_6 & = & \biggl[ \frac{1}{2} \mbox{Tr}(\ep_1 V) \mbox{Tr}(\ep_2 V)
+ \mbox{Tr}(\ep_1 V) (k_2 V \ep_2 V k_2) + (k_2 \ep_1 k_2) 
(k_1 V \ep_2 V k_1) \nonumber \\
& + & \frac{1}{2} (k_2 \ep_1 V k_2)(k_1 \ep_2 V k_1)
+ (k_2 \ep_1 V k_2)(k_1 V \ep_2 k_1) + \frac{1}{2} (k_2 V \ep_1 k_2)
(k_1 V \ep_2 k_1) \nonumber \\ & + & (\mbox{1} \leftrightarrow
\mbox{2}) \biggr] \nonumber \\
d_7 & = & (k_1 \ep_2 k_1) (k_2 \ep_1 k_2) \nonumber \\
d_8 & = & (k_2 \ep_1 k_2) (k_1 \ep_2 V k_1 + k_1 V \ep_2 k_1)
+ (\mbox{1} \leftrightarrow \mbox{2}) \nonumber \\
d_9 & = & (k_2 V \ep_1 V k_2) (k_1 \ep_2 V k_1 + k_1 V \ep_2 k_1)
+ (\mbox{1} \leftrightarrow \mbox{2}) \nonumber \\
d_{10} & = & (k_1 V \ep_2 V k_1) (k_2 V \ep_1 V k_2)~. 
\end{eqnarray}
Our notation is such that all indices
are contracted with the Minkowski metric 
$\eta_{\mu \nu}$, eg. $k_2 \ep_1 V \ep_2 k_1 = k^{\mu}_2 
\ep_{1 \mu \nu} V^{\nu \lambda} \ep_{2 \lambda \rho} k^{\rho}_{1}$.

The scattering amplitude of massless closed string excitations
is now straightforward to evaluate.  One simply substitutes
the ghost (\ref{ghostcorr}) and matter (\ref{masslesscorr})
correlators into the amplitude (\ref{scatamp}) and makes the
change of variables $y = (1-\sqrt{x})/(1 + \sqrt{x})$.  The
resulting integrals are then just beta functions.  In detail one
finds
\begin{eqnarray}
{\cal A}_{D_2}(k_1, \ep_1; k_2, \ep_2) & = & \frac{i}{4} \, g^{2}_c
\, C_{D_2}
\, (2 \pi)^{p+1} \, \delta^{p+1}(k_1 + k_2) \, \Bigl( d_1 \, B(-t/2, 
1+2s) + d_2 \, B(-t/2,2 s) \nonumber \\
& & \hspace{-3em} -  d_3 \, B(1-t/2, 2s) + d_4 \, B(1-t/2,1+2s)
+ d_5 \, B(-1-t/2,1+2s) \nonumber \\
& & \hspace{-3em} +  d_6 \, B(1-t/2, -1+2s) + d_7 \, B(-1-t/2,-1+2s) 
 - d_8 \, B(-t/2,-1+2s) \nonumber \\
& & \hspace{-3em} - d_9 \, B(2-t/2,-1+2s) + d_{10} \, B(3-t/2,-1+2s)
\Bigr) 
\label{exactamp} \\
 & \approx & \frac{i}{4} \, g^{2}_c
\, C_{D_2}
\, (2 \pi)^{p+1} \, \delta^{p+1}(k_1 + k_2) \, \Bigl( -\frac{2}{t}
\, d_1 - \bigl( \frac{2}{t} - \frac{1}{2s} \bigr) \, d_2 - 
\frac{1}{2s} \, d_3 \nonumber \\
& & \hspace{-3em} + 
\bigl(1 + \frac{t}{2} - 2s \bigr) \, d_4 + \bigl(-1 + 4 \frac{s}{t}
+ \frac{t}{2} - 2s \bigr) \, d_5 + \bigl(-1 + \frac{t}{4s}
+\frac{t}{2} - 2s \bigr) \, d_6 \nonumber \\
& & \hspace{-3em} + 2 \bigl(-\frac{2}{t} + \frac{1}{2s}
\bigr) \, d_7 - \bigl( - \frac{2}{t} + \frac{1}{2s} \bigr) \, d_8
+ \frac{1}{2s} \, d_9 - \frac{1}{s} \, d_{10} + {\cal O}(k^4) \Bigr)~,
\label{approxamp}
\end{eqnarray}
where we have recorded both the exact expression for the 
amplitude and the expansion\footnote{The $(\alpha')^0$ part of this
  expression arises from the usual Born-Infeld action, and so should
  match the analogous expression from the Type II superstring
  \cite{GarMye1}. Our expression corrects errors in \cite{GarMye1}
  that are nontrivial for antisymmetric tensor scattering. We thank
  Stefan Theisen and Sebastian Uhlmann for a discussion of this point.}
to quadratic order in momenta
in terms of the $t$-channel momentum transfer\footnote{We
 take $\alpha^{\prime} =2$ for the remainder of
this paper, except when it is restored explicitly.} to the brane defined as $t = - 2 k_1 k_2$ and the $s$-channel 
momentum along the brane defined as $s = - k_1 N k_2$.  The overall
constant $C_{D_2} = e^{-\lambda} C^{X}_{D_2} C^{g}_{D_2}$ 
will be fixed below by comparing to
the field theory limit.

\section{Comparing to the field theory limit}

Given the string scattering amplitudes computed in the last
section, we would now like to find the low energy effective
action for the $Dp$-brane that reproduces them.  For the massless
closed string excitations this is well known to leading
order, ${\cal O}((\alpha^{\prime})^0)$, namely the Born-Infeld
action.  In Einstein frame this is given by
\begin{equation}
S^{(0)}_p = - \tau_p \int d^{p+1} \sigma \, e^{\Phi (-1 - \gamma (p+1)/2)}
\sqrt{-det[\tilde{G}_{ab} + e^{\gamma \Phi} (\tilde{B}_{ab}
+ 2 \pi \alpha^{\prime} F_{ab})]}~,
\label{BI}
\end{equation}
where the induced metric and antisymmetric tensor on the brane
are defined by
\begin{equation}
\tilde{G}_{ab} = G_{MN} \frac{\partial X^{M}}{\partial \sigma^a}
\frac{\partial X^{N}}{\partial \sigma^b}, \,\,\,
\tilde{B}_{ab} = B_{MN} \frac{\partial X^{M}}{\partial \sigma^a}
\frac{\partial X^{N}}{\partial \sigma^b}~,
\end{equation}
and we have defined $\gamma = - 4/(D-2)$.
Our primary interest is to find the field theory terms that 
reproduce the ${\cal O}(\alpha^{\prime})$ terms in the string
scattering amplitude.  Before discussing that calculation however
we shall highlight the analogous ${\cal O}((\alpha^{\prime})^0)$
calculation as a warm-up exercise and to develop some notation.

\FIGURE{\epsfig{file=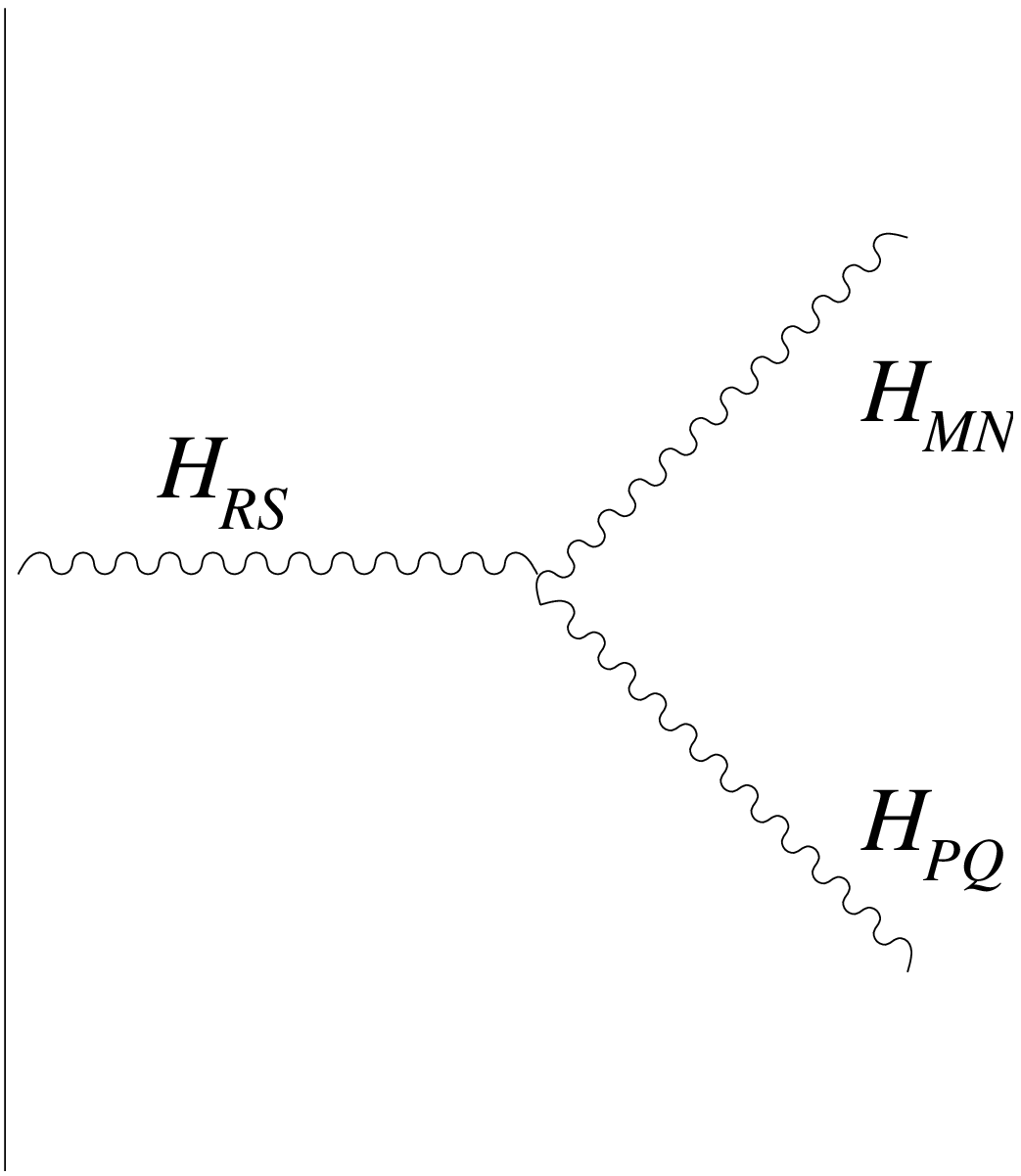,width=5cm}\caption{$t$-channel diagram
for graviton-graviton scattering with graviton internal
leg.}\label{fone}}

\subsection{Graviton scattering at  ${\cal O}((\alpha^{\prime})^0)$}

Here we sketch the comparison of the string amplitude for 
scattering a graviton off a $Dp$-brane to the field theory
amplitude at leading order in $\alpha^{\prime}$.  Aside
from the brane action (\ref{BI}) we shall also need parts
of the bulk low energy effective action for the massless
closed string excitations.  In Einstein frame to zeroth
order in $\alpha^{\prime}$ this is given by
\begin{eqnarray}
S^{(0)}_{bulk} &=& \frac{1}{2 \kappa^2} \int d^{D}x \,
\sqrt{-G} \biggl( R - \frac{1}{12} e^{2 \gamma \, \Phi} H_{MNP} H^{MNP}
\nonumber \\ &+& \gamma \, \partial_{M} \Phi \partial^{M} \Phi \biggr)~,
\label{bulkaction0}
\end{eqnarray}
where our conventions for the Riemann tensor, etc. are
\begin{eqnarray}
R_{MNP}^{\hspace{2.2em} Q} &=&  \partial_{N} \Gamma^{Q}_{MP} - \cdots~,\quad
R_{MP} = R_{MNP}^{\hspace{2.2em} N} ~,\nonumber \\
H_{MNP}  &=&  3 \partial_{[M} B_{NP]}~.
\end{eqnarray}

We now have all the ingredients to check that the Born-Infeld
action (\ref{BI}) correctly reproduces the string amplitude
for scattering of a graviton off a $Dp$-brane to leading
order in $\alpha^{\prime}$.  The relevant field theory diagrams
for this process are displayed in figures \ref{fone}{--}\ref{ffour}.  
There is
a $t$-channel contribution, figure \ref{fone}, coming from the three point
graviton vertex of $R$ in the bulk action (\ref{bulkaction0}), as well
as a three point vertex involving the dilaton, figure \ref{ftwo},
an $s$-channel contribution, figure \ref{fthree},
coming from the exchange of massless
open string fields on the brane, and a contact contribution, figure
\ref{ffour},
 also
coming from the Born-Infeld action (\ref{BI}).

\FIGURE{\epsfig{file=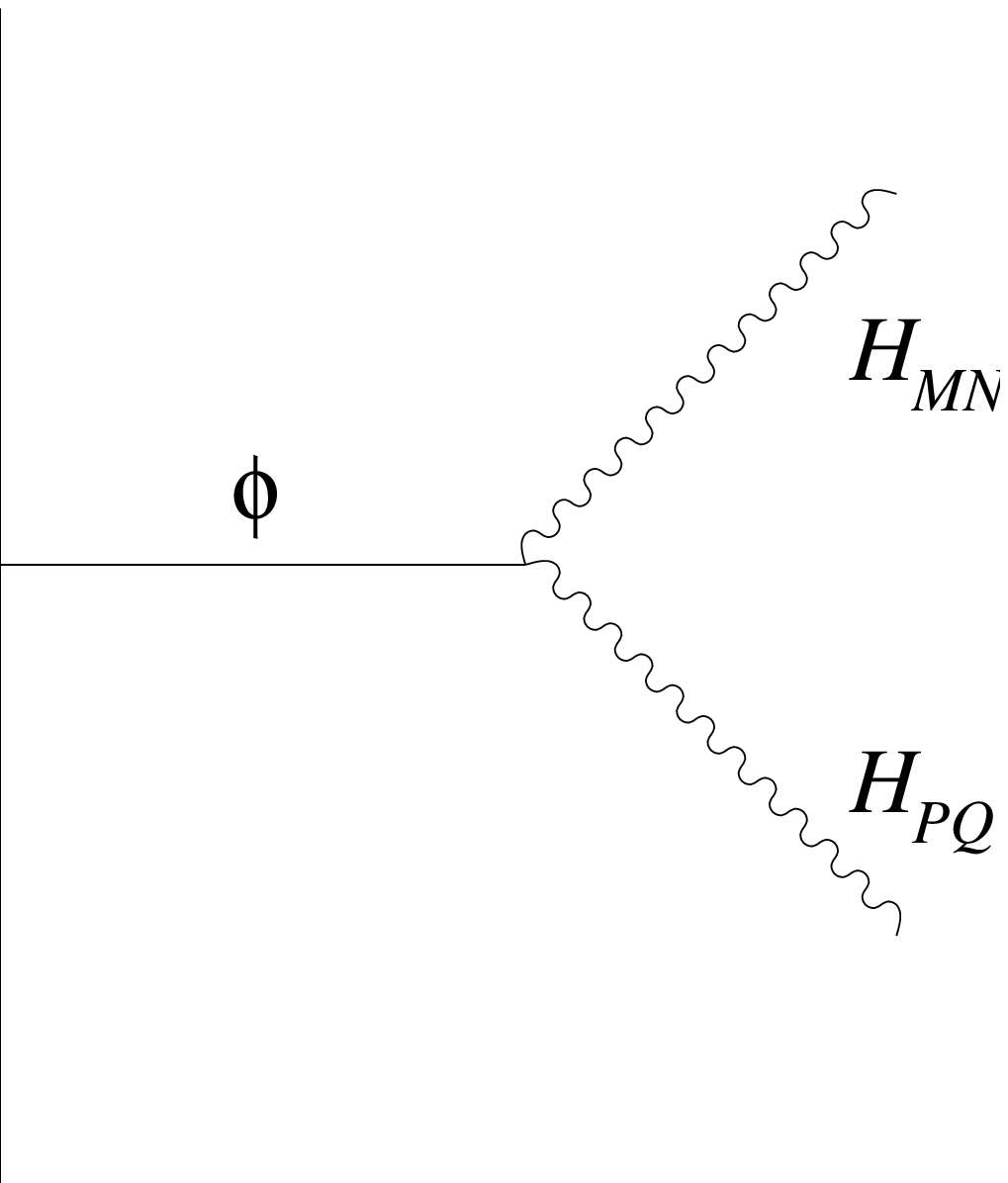,width=5cm}\caption{$t$-channel diagram
for graviton-graviton scattering with dilaton internal
leg.}\label{ftwo}}

To see this in more detail we note first that to compare
to the string theory amplitude we must expand the
bulk (\ref{bulkaction0}) and brane (\ref{BI}) actions about Minkowski
space.  Furthermore we also redefine the remaining massless closed
string fields so as to obtain canonically normalized kinetic
terms as
\begin{eqnarray}
G_{MN} & = & \eta_{MN} + 2 \kappa H_{MN}, \nonumber\\
\Phi & = & \kappa \sqrt{\frac{D-2}{4}} \phi, \nonumber\\
B_{MN} & = & - 2 \kappa b_{MN}~.
\end{eqnarray}
For the bulk action this leads to the usual
momentum space propagators for the graviton and dilaton,
\begin{eqnarray}
\langle \tilde{H}_{MN} \tilde{H}_{PQ} \rangle & = & - \frac{i}{2 k^2}
(\eta_{MP} \eta_{NQ} + \eta_{MQ} \eta_{NP} - \frac{2}{D-2}
\eta_{MN} \eta_{PQ} ) \nonumber \\
\langle \phi \phi \rangle & = & - \frac{i}{k^2}~,
\end{eqnarray}
where we are working in the de Donder gauge, $\partial^{M} H_{MN}
- (1/2) \partial_N H = 0$, to find the graviton propagator.
The three point bulk vertices following from (\ref{bulkaction0})
are similarly straightforward to write down.  The only one
that we shall need for graviton scattering is the three point
graviton vertex coming from the Einstein-Hilbert term.
This can be found in eg. \cite{DeWitt}.

\FIGURE{\epsfig{file=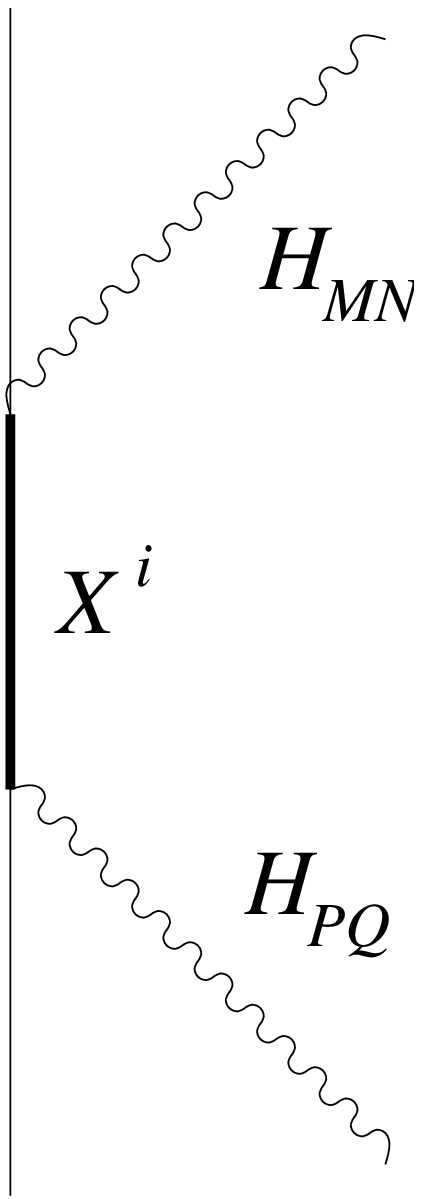,width=3cm}\caption{$s$-channel diagram
for graviton-graviton scattering with $D$-brane
coordinate internal
leg.}\label{fthree}}

Expansion of the Born-Infeld action requires a few steps.
First we fix the worldvolume diffeomorphism invariance
by working in static gauge $X^{a} = \sigma^{a}$.  The
induced metric, for example, expanded about flat space, then
becomes
\begin{equation}
\tilde{G}_{ab} = \eta_{ab} + 2 \kappa H_{ab} + 2 \kappa
\Bigl(\partial_a X^{i} H_{ib}
+ H_{aj} \partial_b X^{j} \Bigr)
+ \partial_a X^{i} \partial_b X^{j} \delta_{ij}
+ 2 \kappa \partial_a X^{i} \partial_b X^{j} H_{ij}~.
\label{inducedG}
\end{equation}
Next we expand the square root and determinant in 
(\ref{BI}) about the induced flat metric appearing
in (\ref{inducedG}).  The massless open string 
fields are furthermore redefined (assuming that
the brane is located at $X^{i} = 0$) as
\begin{eqnarray}
X^{i} & = & \frac{1}{\sqrt{\tau_{p}}} \lambda^{i} 
\label{rescale} \\
A_{a} & = & \frac{1}{2 \pi \alpha^{\prime} \sqrt{\tau_p}} a_{a}~,
\end{eqnarray}
in order to have
canonically normalized kinetic terms.  Finally we note that
the closed string fields evaluated on the brane are
functions of the brane coordinate $X^i$ and should be
Taylor expanded about the background configuration $X^{i} = 0$.
The net result is the expansion of the Born-Infeld action
to quadratic order in fluctuations:
\begin{eqnarray}
S^{(0)}_p & = & - \tau_p \int d^{p+1}x \, \Bigl( \frac{1}{2 \tau_p}
\partial_a \lambda^{i} \partial^a \lambda_i + \frac{1}{4 \tau_p}
f_{ab} f^{ab} + \kappa \, \frac{p-D/2 + 2}{\sqrt{D-2}} \, \phi + \kappa 
H_{a}^{\hspace{0.5 em} a} \nonumber \\
& & \hspace{-3em} + \frac{\kappa}{\sqrt{\tau_p}} 
(\lambda^{i} \partial_i H^{a}_{\hspace{0.5em}a} + 2 \partial^a
\lambda^{i} H_{ia} ) + \kappa \, \frac{p-D/2 +2}{\sqrt{D-2}} \, \lambda^i
\partial_i \phi + \frac{\kappa}{\sqrt{\tau_p}} b_{ab} f^{ab}  \nonumber \\ 
& & \hspace{-3em} -  \kappa^2 (H^{a}_{\hspace{0.5em} b}
H^{b}_{\hspace{0.5em} a} - \frac{1}{2} H^{a}_{\hspace{0.5em} a}
H^{b}_{\hspace{0.5em} b})
+ \kappa^2 \frac{p- D/2 + 2}{\sqrt{D-2}} \, \phi
H^{a}_{\hspace{0.5em} a} + \kappa^2 b_{ab} b^{ab} + \cdots \Bigr)~. 
\label{BIexpanded}
\end{eqnarray} 
The momentum space $\lambda^i$ and $a_a$ propagators following from this
action (working in Lorentz gauge $\partial^a a_a = 0$ for
the vector) are
\begin{eqnarray}
\langle \tilde{\lambda}^i \tilde{\lambda}^j \rangle = - 
\frac{i}{k^2} \delta^{ij},
\nonumber \\
\langle \tilde{a}_a \tilde{a}_b \rangle = - \frac{i}{k^2} \eta_{ab}~.
\end{eqnarray}

Now it is a simple matter to evaluate the amplitudes associated
with the Feynman diagrams pictured in figures 1{--}4.
The $t$-channel, $s$-channel, and contact term contributions
are given respectively by
\begin{eqnarray}
{\cal A}_{t}(h,h) & = & i \frac{\tau_p \kappa^2}{t} \biggl( - 
\Bigl(\frac{t}{2}
- s \Bigr) \mbox{Tr}(\ep_1
\ep_2) + \frac{t}{2} \mbox{Tr}(\ep_1 V \ep_2) - 
\mbox{Tr}(\ep_1 V) \, (k_1 \ep_2 k_1) \nonumber \\
& - & 2 \, k_1 \ep_2 V \ep_1 k_2
+ k_1 \ep_2 \ep_1 k_2 - 4 \, k_1 V \ep_1 \ep_2 k_1 +
(1 \leftrightarrow 2) \biggr) \nonumber \\
{\cal A}_{s}(h,h) & = & i \frac{\tau_p \kappa^2}{4s} 
\biggl( \frac{1}{2} \Bigl(\frac{t}{2}
- s \Bigr) \mbox{Tr}(\ep_1 V) \mbox{Tr}(\ep_2 V)
+ \mbox{Tr}(\ep_1 V) (k_1 \ep_2 k_1 - k_1 V \ep_2 V k_1) \nonumber \\
& - & \, k_1 V \ep_1 \ep_2 V k_2
 -  \, k_1 V \ep_1 V \ep_2 V k_2 
+(1 \leftrightarrow 2) \biggr) \nonumber \\
{\cal A}_{contact}(h,h) & = & i \tau_p \kappa^2 \Bigl( 2 \mbox{Tr}
(\ep_1 N \ep_2 N) - \mbox{Tr}(\ep_1 N) \mbox{Tr}(\ep_2 N) \Bigr)~.
\end{eqnarray}
Summing up these amplitudes one can show
after some rearranging that this indeed agrees
with the string amplitude (\ref{approxamp}) to zeroth order in 
$\alpha^{\prime}$ if one makes the identification
\begin{equation}
\frac{1}{4} g^{2}_c C_{D_2} (2 \pi)^{p+1} = \frac{1}{2} \tau_p
\kappa^2~.
\label{fixcoeff}
\end{equation}

\subsection{Graviton scattering at  ${\cal O}(\alpha^{\prime})$}

Now we would like to construct the piece of the low energy effective
action that accounts for the ${\cal O}(\alpha^{\prime})$ part of
the string amplitude for scattering a graviton from a $Dp$-brane.
The computation is exactly analogous to that presented in the
previous section for the ${\cal O}((\alpha^{\prime}))$ terms.
There will be $t$-channel, $s$-channel, and contact term 
contributions represented by the Feynman diagrams of 
figures 1{--}4.  The $t$-channel contributions come
from terms in the effective action that are already known, so
we begin with these.

The $t$-channel contributions at quadratic order in the
momenta come from bulk 3-point vertices which are quartic
in derivatives and contain either three gravitons or
two gravitons and a dilaton (since only the graviton and
dilaton have tadpoles on the brane).  Such terms have already 
been worked out in \cite{Zwe,MetTse1} and are given by
\begin{equation}
S^{(1)}_{R^2,bulk} = \frac{1}{2 \kappa^2} \int d^{D}x \, \sqrt{-G}
\frac{\alpha^{\prime}}{4} e^{\gamma \Phi} (R^{MNPQ} R_{MNPQ}
- 4 R^{MN} R_{MN} + R^2)~.
\label{RRterms}
\end{equation}
Actually only the coefficient of the square of the Riemann tensor
is fixed by the string S-matrix, while the coefficients of
the Ricci squared terms are arbitrary.  They can be fixed by
a local field redefinition (which leaves the perturbative S-matrix
invariant) to the values given above.  The above choice, 
corresponding to the Gauss-Bonnet combination, however
is very convenient because the quadratic term in $H_{MN}$
appearing in it's expansion is a total derivative, and
therefore there is no graviton propagator correction arising
from this combination.  For a detailed discussion of
the ambiguities in the string effective action see
\cite{Tse1,MetTse1,MetTse2}.

Given this choice for the higher derivative terms in (\ref{RRterms})
we compute the $t$-channel
amplitudes for graviton or dilaton exchange.  The net result is
given by
\begin{eqnarray}
{\cal A}_{t,R^2}(h,h) & = & i \frac{\alpha^{\prime}}{2}
\tau_p \kappa^2 \biggl( \Bigl(\frac{t}{2} - 2s \Bigr) \mbox{Tr}(\ep_1
\ep_2) - t \mbox{Tr}(\ep_1 \ep_2 N) + 2 \Bigl(1 - 2 \frac{s}{t} \Bigr)
(k_1 \ep_2 \ep_1 k_2) \nonumber \\
& - & \frac{4}{t} ((k_1 \ep_2 k_1)(k_2 N \ep_1 k_2)
+(k_2 \ep_1 k_2)(k_1 N \ep_2 k_1)) + \frac{2}{t} (k_1 \ep_2 k_1)
(k_2 \ep_1 k_2) \nonumber \\
& - & 2(k_1 N \ep_2 \ep_1 k_2 + k_1 \ep_2 n \ep_1 k_2
+ k_1 \ep_2 \ep_1 N k_2) \biggr)~.
\end{eqnarray}
This amplitude by itself already provides a check on our
calculations.  The only way in which to get terms proportional
to $1/t$ is via such $t$-channel amplitudes.  Since the
overall coefficient of the string amplitude (\ref{exactamp}) was
already fixed in (\ref{fixcoeff}) we see that all the $1/t$ terms
in the above amplitude give us non-trivial consistency
checks, which fortunately work out.

\FIGURE{\epsfig{file=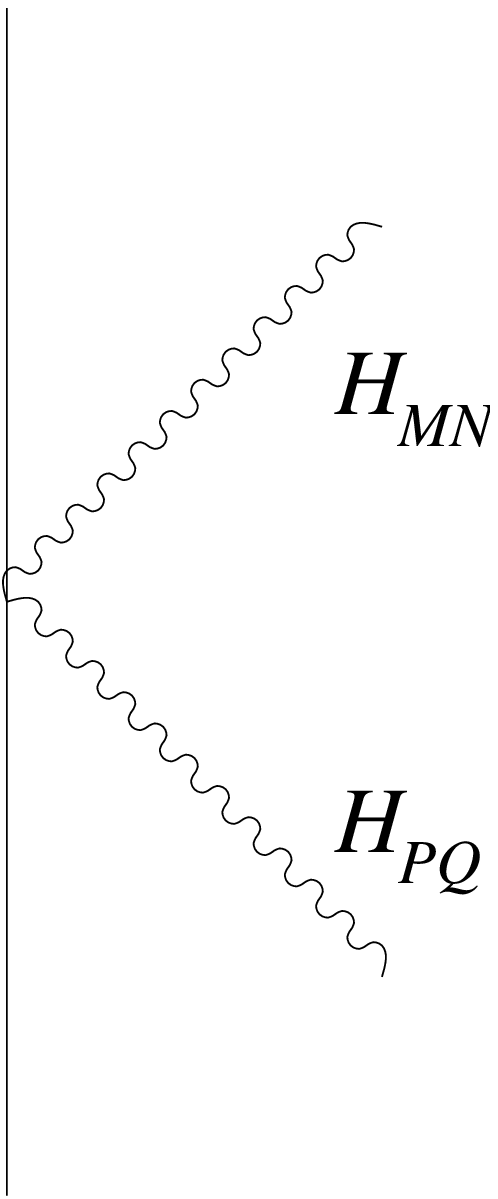,width=3cm}\caption{Contact term diagram
for graviton-graviton scattering.}\label{ffour}}

For the $s$-channel and contact term contributions we need to
work a little bit harder.  The Born-Infeld action clearly
cannot account for these terms by itself (that is, it cannot account
for the remaining terms in the string amplitude) because
it gives no contribution to graviton scattering at 
${\cal O}(\alpha^{\prime})$.  Therefore we are forced to
add new terms to the brane action.  The possible terms that
can contribute to the amplitude at ${\cal O}(\alpha^{\prime})$
come in three forms: (1) linear in $H_{MN}$ and quadratic
in derivatives, (2) linear in both $H_{MN}$ and $X^{P}$ and
cubic in derivatives, and (3) quadratic in both $H_{MN}$ and
derivatives, all constrained of course by the symmetry requirement
of diffeomorphism invariance.  A term of type (1) will necessarily give rise
to a term of type (2) upon expanding about the background brane
configuration, but not vice versa.  (1) and (2) give rise
to $s$-channel contributions and (3) to contact term contributions.
The most obvious term to add is the induced Ricci
scalar on the brane; however, there are in fact five
possible terms which satisfy this criterion.  In terms
of the brane action they are given by
\begin{eqnarray}
S^{(1)}_{brane} & = & - \frac{1}{2 \kappa_{p}^2} \int d^{p+1}x
\, \sqrt{- \tilde{G}} ( \tilde{R} + \beta_1 K^{i}_{ab}
K^{ab}_{i} + \beta_2 K^{ia}_a K^{b}_{ib} + \beta_3
\perp^{MN} R_{MN} \nonumber \\
& + & \beta_4 \perp^{MN} \perp^{PQ} R_{MPNQ} )~,
\label{Rbrane}
\end{eqnarray}
where some explanation is required.  We use $\tilde{} \,$'s to denote
induced quantities, so eg. $\tilde{R}$ is the Ricci scalar 
computed from the induced metric $\tilde{G}_{ab}$.  $K^{i}_{ab}$
is the second fundamental form whose definition
and explicit expansion we give in the appendix.  The
key point at the moment is that it contains terms linear
in derivatives and linear in the graviton.  The remaining
terms are transverse projections of the bulk Riemann tensor
evaluated on the brane.  The projector $\perp_{MN}$ is
defined in terms of the induced metric as
\begin{equation}
\perp^{MN} = G^{MN} - \tilde{G}^{MN}~,
\label{perp}
\end{equation}
where
\begin{equation}
\tilde{G}^{MN} = \partial_{a} X^{M} \tilde{G}^{ab} \partial_b X^{N}~.
\end{equation}
These are the only\footnote{Another possible term would be the
Ricci scalar $R$ evaluated on the brane.  This term however
is not independent as follows from the generalized Gauss identity.} 
possible terms consistent with
diffeomorphism invariance (both bulk and worldbrane) and our
constraint that they contribute to the contact term amplitudes
at this order.

Now we fix the coefficients in (\ref{Rbrane}) by computing all
$s$-channel and contact term diagrams from the brane (\ref{BI},\ref{Rbrane})
action and comparing to the string amplitude.
Note that the expansion of the second fundamental
form as given in the appendix contains the contribution
$K_{ab}^{i} = \partial_a \partial_b X^i + \cdots$.  From the
brane action (\ref{Rbrane}) this would imply for 
$\beta_2 \neq - \beta_1$ that there is a quadratic
term for the $X^i$'s proportional to $(\beta_1 + \beta_2)
(\partial^a \partial_a X^i)^2$.  Such a term would
modify the propagator of the $X^i$'s (or rather $\lambda^i$'s
after the rescaling in (\ref{rescale})).  In section
\ref{fieldredef} we discuss the field redefinition
ambiguities associated with the $X^i$'s and
show that one can use the field redefinition freedom
to choose $\beta_2 = - \beta_1$.  Using the 
standard propagator for the $\lambda^i$'s is therefore
equivalent to using the field redefinition freedom
to fixing the above relation between $\beta_1$ and
$\beta_2$.
 
Another point to note is that while the Born-Infeld action by itself does not
contribute to the scattering amplitude at this order,
it will contribute to the $s$-channel diagrams when 
combined with (\ref{Rbrane}).  Specifically the $\lambda
\partial H$ type terms from (\ref{BIexpanded}) combine
with the $\partial^2 \lambda \partial H$ type terms
from the expansion of the second fundamental form
terms in (\ref{Rbrane}).  A lengthy calculation
of the amplitudes corresponding to the diagrams
displayed in figures 1{--}4, leads to a lengthy
expression for the sum of the $s$-channel and 
contact term contributions at ${\cal O}(\alpha^{\prime})$.  
The upshot is that
in comparing the entire field theory amplitude
(including the $t$-channel contribution as well) with the string
amplitude leads to 19 equations (not including the
trivial equations following from the $1 \leftrightarrow 2$
exchange symmetry of the amplitude) for the 5 coefficients.
This obviously fixes the coefficients and gives many
consistency checks, all of which work out fortunately.
In the end we find that the coefficients (in units where
$\alpha' = 2$) are given by
\begin{equation}
\frac{1}{\kappa^{2}_p} = 2 \tau_p, \,\,\, \beta_1 = -1, \,\,\,
\beta_2 = 1, \,\,\, \beta_3 = 0 = \beta_4 ~.
\label{hhcoeff}
\end{equation}

\subsection{Dilaton and Antisymmetric Tensor scattering}

The analysis for the dilaton and antisymmetric tensor is
completely analogous.  The leading order analysis in
particular just amounts to checking that the Born-Infeld
correctly reproduces the string scattering amplitudes
to zeroth order in $\alpha^{\prime}$.

At ${\cal O}(\alpha^{\prime})$ we need to add new
terms to the brane action, just as for the graviton,
in order to account for the string amplitude.  For
the dilaton the analysis is somewhat simplified in that
there is no field theory $t$-channel contribution 
at this order as there is no bulk vertex which is
quartic in derivatives and either cubic in $\phi$'s
or quadratic in $\phi$'s and linear in the graviton,
as was shown\footnote{More precisely,
\cite{MetTse2} showed that one can use the field
redefinition freedom that leaves the S-matrix invariant
to remove any such terms.} in \cite{MetTse2}.  The possible
terms that one could add to the brane action which are
quadratic in derivatives are given by\footnote{
There is one other term that one should add to this action
given by $K_{a}^{ai} n_{i}^{M} \nabla_M \Phi$.  In section
\ref{fieldredef} we show that such a term can be removed
by field redefinitions that leave the S-matrix invariant.}

\begin{eqnarray}
S^{(1)}_{p,\Phi} & = & - \tau_p \int d^{p+1}x \, \Bigl(\gamma_1
\perp^{MN}(1 -(1+\gamma (p+1)/2) \Phi ) \nabla_M \partial_N \Phi
+ \gamma_2 \tilde{G}^{ab} \partial_a \Phi \partial_b \Phi
\nonumber \\
& + & \gamma_3 \perp^{MN} \partial_M \Phi \partial_N \Phi \Bigr)~.
\label{Phibrane}
\end{eqnarray}
Note that the relative coefficient between the two $\gamma_1$
terms is fixed because an undifferentiated $\Phi$ can
only appear in the action through the expansion of an
exponential.  In particular in the string frame the
brane action corresponding to tree level string processes
is of the form $e^{-\Phi} {\cal L}(G,B,\nabla \Phi,...)$.
Transforming to Einstein frame one gets an overall factor
of $\exp (- (1 + \gamma (p+1)/2) \Phi)$ multiplying the $\perp^{MN}
\nabla_{M} \partial_N \Phi$ term, which after expanding results
in the above relative coefficient.

If one now computes the total amplitude for dilaton scattering
off the brane to quadratic order in momenta (consisting of only
the $s$-channel and contact term contributions), one finds that
$\gamma_1$ remains a free parameter.  In order to fix this
freedom one must compute the amplitude for converting a graviton
into a dilaton.  The string amplitude again follows from
(\ref{scatamp}) by substitution of the appropriate polarization
tensors.  The corresponding computation on the low-energy effective
action side reveals that no additional terms on top of those
already added in the previous section must be added to account
for the string amplitude at ${\cal O}(\alpha^{\prime})$.
In particular the argument in the previous paragraph
fixing the overall dilaton
dependence of the brane action predicts the necessary (though
not sufficient)
$\Phi-H_{MN}$ coupling from the induced Ricci scalar term
found in the last section to correctly reproduce the string
amplitude.  In particular this result tells us that $\gamma_1=0$.
The remaining $\gamma_i$ coefficients are then fixed to 
be 
\begin{equation}
\gamma_2 = -\gamma_3 = - \frac{\gamma^{2}}{2} 
\Bigl((p-D/2 +2)^2 - 2 (p-D/2 +2) + D-2 \Bigr)~.
\end{equation}

The analysis for the  antisymmetric tensor is similar to that
of the graviton so we shall skip the details and highlight
the main points of the ${\cal O}(\alpha^{\prime})$ calculation.  
The field theory amplitude for scattering
the antisymmetric tensor off a $Dp$-brane again contains a
$t$-channel contribution.  At ${\cal O}(\alpha^{\prime})$ the
relevant bulk
3-point vertex arises from the following term 
worked out in \cite{MetTse2},
\begin{equation}
S^{(1)}_{bulk,H^2} = -\frac{1}{16 \,\kappa^2} \int d^{D}x \, \sqrt{G}
e^{3 \gamma \Phi} R^{MNPQ} H_{MNR} H_{PQ}^{\hspace{1.2em} R}~.
\end{equation}
The Born-Infeld action by itself makes no other contributions
to the scattering amplitude at this order, so as 
before we must add terms to the 
brane action which are either linear in $B_{MN}$ and
quadratic in derivatives, or quadratic both in derivatives
and in $B_{MN}$.  The possible terms are 
\begin{eqnarray}
S^{(1)}_{p,B} & = & - \tau_p \int d^{p+1}x \, \Bigl( \rho_1
\tilde{H}_{abc} \tilde{H}^{abc} + \rho_2 \perp^{MN} H_{MPR}
H_{N}^{\hspace{0.7em} PQ} + \rho_3 \perp^{MN} \perp^{PQ} H_{MPR}
H_{NQ}^{\hspace{1.2em} R} \nonumber \\
& + & \rho_4 \perp^{MN} \perp^{PQ}
\perp^{RS} H_{MPR} H_{NQS} + \rho_5 (\tilde{B}^{ab} 
+ 2 \pi \alpha^{\prime} F^{ab}) \nabla^{M} H_{Mab} \Bigr)~.
\label{Bbrane}
\end{eqnarray}
Working through the details of computing the $s$-channel and
contact term contributions and comparing to the string 
amplitude results in the coefficients
\begin{eqnarray}
\rho_1 & = & \frac{1}{12} \Bigl(1 - \frac{p+1}{D-2} \Bigr),
\,\,\,\,\,\,\,\,\,\,
\rho_2 = \frac{1}{4} \Bigl(1 - \frac{p+1}{D-2} \Bigr) \nonumber \\
\rho_3 & = & - \frac{1}{2} \Bigl( 1 - \frac{1}{2} \frac{p+1}{D-2} 
\Bigr), \,\,\, \rho_4 = \frac{1}{4} \Bigl(1 - \frac{1}{3}
\frac{p+1}{D-2} \Bigr)~,
\end{eqnarray}
while $\rho_5$ cannot be determined at this order because it
gives vanishing contribution to this particular scattering 
amplitude.

\section{Field redefinition ambiguities}
\label{fieldredef}

We have mentioned in passing in the previous sections that
the low energy effective action is ambiguous due to the
freedom of making local field redefinitions 
\cite{MetTse1,MetTse2,Tse1}.  Namely
local field redefinitions that preserve the required
symmetries of the action and that do not change the pole
structure, leave the S-matrix invariant.  We would now
like to investigate to what extent the results we have
obtained for the ${\cal O}(\alpha^{\prime})$ terms in 
the brane action are fixed independently of field
redefinitions.

The most general local field redefinition of the massless closed
string fields that preserves diffeomorphism invariance and the
property that the dilaton cannot appear undifferentiated in the
action (except in the argument of an overall exponential) is given
by \cite{MetTse2}
\begin{eqnarray}
\delta G_{MN} & = & \alpha^{\prime} \Bigl( b_1 R_{MN} + b_2 \partial_M
\Phi \partial_N \Phi + e^{3 \gamma \Phi} b_6 H_{MPQ} H_{N}^{\hspace{0.5em}
PQ} + G_{MN} \bigl(b_3 R + b_4 (\partial \Phi)^2 \nonumber \\
& + & b_5 \nabla^M \partial_M
\Phi + b_7 e^{3 \gamma \Phi} H^2 \bigr) \Bigr)
\nonumber \\
\delta \Phi & = & \alpha^{\prime} \Bigl( c_1 R + c_2 (\partial \Phi)^2
+ c_3 \nabla^M \partial_M \Phi + c_4 e^{3 \gamma \Phi} H^2 \Bigr)
\nonumber \\
\delta B_{MN} & = & \alpha^{\prime} e^{\gamma \Phi} \Bigl( d_1 \nabla^R
H_{MNR} + d_2 H_{RMN} \nabla^R \Phi \Bigr)~,
\label{deltaB}
\end{eqnarray}
where $\gamma = -4/(D-2)$.
Such a field redefinition was already used to fix coefficients of
some of the terms in the bulk action \cite{MetTse1,MetTse2}.
A different choice of these coefficients would in general change
the $t$-channel contribution to the amplitudes computed previously,
and therefore the coefficients of the brane terms that we have
added.  To see exactly which brane action terms are affected by
these field redefinitions we need only vary the brane action
with respect to (\ref{deltaB}).  One sees 
immediately that the Born-Infeld part of the action will
be unaffected as it is ${\cal O}((\alpha^{\prime})^0)$ in the
closed string fields while the field redefinitions 
(\ref{deltaB}) are 
${\cal O}(\alpha^{\prime})$.  Variation
of the Born-Infeld action with respect to the field
redefinitions (\ref{deltaB}) however will give
to rise to changes in
the ${\cal O}(\alpha^{\prime})$ terms computed in the
previous sections.  

As a specific example, the gravitational terms that
are produced from varying the Born-Infeld are
\begin{eqnarray}
\delta (B.I.) & = & \alpha^{\prime} (B.I.) \biggl( \Bigl(\frac{b_1}{2}
+\frac{p+1}{2} b_3 -(1+\gamma (p+1)/2) c_1 \Bigr) (\tilde{R}
+ K^{i}_{ab} K^{ab}_{i} - K^{ai}_a K^{b}_{bi} \nonumber \\
& & \hspace{-4em} - \perp^{MN} \perp^{PQ} R_{MPNQ} )
+ \Bigl(\frac{b_1}{2}
+ (p+1) b_3 -2 (1+\gamma (p+1)/2) c_1 \Bigr) \perp^{MN} R_{MN} \biggr)
\end{eqnarray}
after some rewriting.  The first point to note is that
only two of the three coefficients $b_1$, $b_3$, and $c_1$
are independent.  This means that only two of the five possible
terms that we considered in (\ref{Rbrane}) can be set to zero
by a field redefinition.  In the `scheme' in which we have
used so far we have set the coefficients of
the $\perp^{MN} R_{MN}$ and $\perp^{MN} \perp^{PQ} R_{MPNQ}$
terms to zero.

In an analogous way one can show that all the ${\cal O}(\alpha^{\prime})$
terms in the dilaton considered in (\ref{Phibrane}) can be removed
by a field redefinition.  The necessary field redefinition however
would introduce $p$ (one less the brane dimension) 
dependence into the bulk action.  Such a dependence
would be a bit strange as we have no reason to expect any
particular $p$ to appear.  Furthermore if we have two
or more branes with at least two different brane dimensions,
then one could simultaneously remove the ${\cal O}(\alpha^{\prime})$
terms for only those branes of a fixed dimension, and 
not the others.

The same type of analysis for the antisymmetric tensor
shows that of the four coefficients $b_6$, $b_7$, $c_4$,
and $d_1$, only three are independent.  In fact $d_1$
corresponds to changing the value of $\rho_5$ which 
we were unable to fix from our analysis.  Two of the
remaining coefficients can then be used to remove two
more ${\cal O}(\alpha^{\prime})$ terms in (\ref{Bbrane}).

As for the closed string fields one may consider field redefinitions
of the open string fields that leave the S-matrix invariant.
In fact we have already used this freedom in an essential way
in obtaining the propagator for the $\lambda^i$'s.  As before
the idea is to consider the most general field redefinition that
preserves the symmetries and that does not change the pole structure
of the elementary fields.  As an example consider the $X^i$ field
redefinition
\begin{equation}
\delta X^i = e_1 K_{a}^{ai} + e_2 \partial^i \Phi~.
\end{equation}
This is the most general field redefinition that is linear
in derivatives (one could of course consider higher derivative
terms, but these would not effect the terms that are of interest
to us).  Acting on the Born-Infeld action this would generate
the terms
\begin{equation}
\delta ( B.I.) = - \tau_p \int d^{p+1}x \, e^{\beta \Phi} 
\sqrt{- \tilde{G}} \Bigl( - e_1 K_{a}^{ai} K^{b}_{bi}
+(\beta e_1 - e_2) K_{a}^{ai} \partial_i \Phi +
\beta e_2 (\partial_i \Phi)^2 \Bigr)~,
\end{equation}
where $\beta = -(1 + \gamma (p+1)/2)$.
Two of the above three terms therefore have ambiguous
coefficients.  As we commented on previously, we used this
field redefinition freedom to set the coefficient of
the $K_{a}^{ai} \partial_i \Phi$ term to zero and the
coefficient of the $K_{a}^{ai} K^{b}_{bi}$ term, denoted
$\beta_2$ in (\ref{Rbrane}), equal to minus the coefficient
of the $K_{ab}^i K^{ab}_i$ term, i.e., to set $\beta_2 = - \beta_1$.
Note that this field redefinition, combined with the
above closed string field redefinition, would not allow one to remove
all the induced gravitational terms on the brane that we have
found.

\section{Scattering of brane-localized fields and the Einstein-Hilbert
  term } 

 The Einstein term  and the extrinsic curvature 
 terms in (\ref{Rbrane})  on the brane were derived 
 by analyzing the scattering of gravitons 
 off the brane. They also affect the 
 scattering of particles living on the brane. 
 We may consider, for example, the $ 2 \rightarrow 2 $ 
 scattering of scalars $X^i$ associated to the transverse 
 fluctuations. For such a scattering process, 
 a similar analysis to the one  of  previous sections, 
 can be performed. There is a string answer, which can be 
 computed in string perturbation theory. The leading answer
 comes from a disc diagram with four open string 
 vertex operators. The next diagram is a cylinder with the same 
 vertex operators. The amplitude can be compared to field theory 
 with a brane action coupled to the bulk action. Further interactions 
 in the brane action can be deduced by requiring agreement 
 with the string S-matrix. 
 
 Consider the scattering of $\langle X(k_1)   X(k_2)  X(k_3)  X(k_4) \rangle $, 
 with $1$ and $2$ being the incoming particles, and $3$ and $4$ 
 being the outgoing particles. 
 The leading t-channel diagram involves 
 exchange of a graviton which propagates in the bulk. 
 There are corrections to the $\langle XXh \rangle $ vertex coming from 
 the new gravitational terms we have derived. These corrections 
 affect the leading order in $g_s$  amplitude, but at higher 
 orders in momenta.  The new terms also affect the
 next to leading order in the string expansion, by changing 
 the propagator. It is instructive to consider the form of this 
 correction to the four point amplitude. 

 The leading contribution in the t-channel 
 gives
\begin{equation}  
k_1.k_3 ~k_2.k_4 P(\zeta ) \int d^{D-p-1}p~
  \frac{ e^{i {\vec p} \cdot {\vec  \epsilon } }}
{  q^2 + p^2 }  ~\delta^{p+1} \left( \sum_i k_i\right)
   ~.
\label{propone}
\end{equation} 
  The momenta $k_i$ have components parallel to the brane only, 
 and $P$ is a polarization dependent factor. The momenta $k_i$ 
 satisfy the on-shell condition $k_i^2 = 0$. 
 The momentum of the intermediate graviton is 
 $(q+p) $, 
 with $q = k_1 + k_3 $ by momentum
 conservation along the brane, and $p$ is the  component 
 perpendicular to the brane. The factor $e^{i q.\epsilon}$ 
involves a vector in the transverse directions,  $ \vec \epsilon$
which is a point splitting regulator for the integral 
that separates different brane-localized vertices.
We can extract a long distance potential from 
 (\ref{propone}) by performing the integral over $p$ and Fourier
 transforming with respect to the spatial components of $q$.
 This leads to a potential 
 ${ \frac{1}{  r^{D-3}}} $ as expected for gravity in 
 $D$ dimensions.

 The next order contribution is of the form :  
\begin{equation} 
g_sl_s^{D-p-1} ~k_1.k_3~ k_2.k_4~\tilde P( \zeta )  \int d^{D-p-1}p_1~ \int d^{D-p-1}p_2~ 
  \frac{ q^2 e^{i ( {\vec p_1} \cdot {\vec \epsilon_1}
+ {\vec p_2} \cdot {\vec \epsilon_2} )} }{ (  q^2 + p_1^2)
   (  q^2 + p_2^2  ) }  ~ 
 \delta^{p+1} \left( \sum_i k_i\right) ~.
\label{secprop} 
\end{equation}  
 The power of $l_s$ is easily understood
 from dimensional analysis.
 Performing the integral over the transverse momenta, 
 we get a correction which behaves lie $q^{2D-2p-2} $.
 A spatial Fourier transform gives a potential 
 $ g_s l_s^{D-p-1} /  r^{2D-p-4} $.   
 In the case $D=5,p=3$, this form was derived as 
 a limit of an exact propagator in \cite{Dvali}. 

 The terms in (\ref{secprop}) and (\ref{propone}) 
 become comparable when the momenta exchanged are of 
 order $1/g_s^{ 1/(D-p-1)} l_s $, corresponding to 
 a distance scale of $ g_s^{ 1/(D-p-1) } l_s $. However our low energy
 effective description of the scattering breaks down once the momentum
 become of order $1/l_s$.
 This prevents us from seeing 
 a transition between the D-dimensional gravitational 
 force law to a $(p+1)$-dimensional force law in the context of
 perturbative string theory.

However it is interesting to try to extrapolate our results to
strong coupling to make the brane Einstein term dominate over the
bulk Einstein term. Of course the detailed results cannot be trusted
in this regime, but we can hope to learn something about the
qualitative behavior. The relevant length scales are:
\begin{eqnarray}
& & l_s \quad\qquad\quad{\rm string ~scale} \nonumber \\
 l_{bulk}&=&  g_s^\frac{2}{D-2} l_s\quad{\rm bulk ~Planck ~scale}\nonumber \\
 l_{brane}&=& g_s^\frac{1}{p-1} l_s\quad{\rm brane ~Planck ~scale}\nonumber \\
 l_t&=& g_s^\frac{1}{D-p-1}l_s\quad{\rm transition ~scale} ~.
\end{eqnarray}
In particular, we should require that the longitudinal length scale
$L$ be larger than $l_s$, $l_{brane}$, but be smaller than $l_t$. This
can only be satisfied for $g_s>1$. These inequalities also lead to the
condition $p>D/2$. The marginal case $p=D/2$ suggests that in the
context of ten-dimensional string theory, the five-brane may be the
most likely candidate for brane localized gravity. It would be very
interesting to develop a calculational scheme for computing the low
energy effective action in this case.
The D5-brane has also made an appearance (although in a somewhat
different manner than considered here) in the context of localizing
gravity in \cite{KarRan1,KarRan2}.
Superstring embeddings of the scenario of \cite{Dvali,Dvali2}
have been recently discussed in \cite{Kir}.

\section{ Comments on tachyon terms on the brane }

It has been implicit in the above analysis that we are taking a low
momentum limit, retaining only the massless fields in the bulk, and on
the worldvolume of the brane. One might worry this is not a sensible
limit to take in a theory with a tachyon (both in the bulk and on the
brane). 
Starting from string field
theory, we can take such a limit provided we redefine the massive
fields (including the tachyon) to absorb terms linear in the massive
fields \cite{Tse1}. For example, a tachyon coupling of the form $T X^M
X_M$ leads to a tachyon equation of motion
\begin{equation}
\partial^2 T - m_T^2 T = X^M X_M~,
\end{equation}
and $T$ cannot be integrated out directly.
With a shift of the tachyon field,
\begin{equation}
\tilde T = T - \frac{ X^M X_M}{\partial^2  - m_T^2 }
\end{equation}
we can make a systematic low momentum expansion about $\tilde T=0$.

However, it is interesting to consider the
general properties of the coupling of the open string tachyon to the
massless fields \cite{sen,sen2}, despite the fact that such an
expansion about low momentum is not systematic.
In general, all the brane-localized couplings of the 
 bulk fields will be functions of the open string tachyon. 
 For example, at lowest order in derivatives, the Einstein term on the brane 
 is multiplied by a function $f(T_0)$. We have 
 determined, in an expansion in powers of $T_0$, the leading 
 term in $\int f(T_0) R \sqrt{g } $. 
 Higher terms can be determined by further comparison 
 of S-Matrix calculations and brane action. One general property 
 of functions like $f(T_0) $ follows immediately 
 from Sen's conjectures \cite{sen}.
 According to these conjectures, the minimum 
 of the tachyon potential describes the 
 vacuum configuration without the brane. 
 The absence  of the brane implies that {\it all } 
 brane localized couplings should vanish at the minimum 
 of the potential. For example.  the value of $f$ when $T_0$ 
is at the minimum 
 of the open string tachyon potential should be zero. 
 A simple way to ensure this would be to have all the 
 brane-localized terms to be multiplied by the one function 
 which is known to have this property, namely the tachyon 
 potential itself. Another possibility is to have these 
 functions be polynomials in the potential. 
 We hope to report on further S-matrix calculations 
 exploring these possibilities in the near future.

\acknowledgments

We would like to thank Brandon Carter, Gia Dvali, Antal Jevicki and
Stefan Theisen for
helpful conversations. D.L. wishes to thank the AEI Potsdam, Germany
for hospitality during the completion of this work.
This research is supported in part by DOE grant DE-FE0291ER40688-Task A.

\appendix
\section{The first and second fundamental tensors}

In this appendix we summarize some useful, but perhaps less
well known, facts about submanifolds.  A more detailed
account of the information given here can be found
in \cite{Car}.  Let $\Sigma$ be a 
$p$-dimensional submanifold of a $D$-dimensional
semi-Riemannian space ${\cal M}$ with metric $G_{MN}$.
In some local coordinate system $X^{M}$ let
$\Sigma$ be parameterized by the world-volume coordinates
$\sigma^{a}$ with embedding given by $X^{M}(\sigma^a)$.
$\Sigma$ then has an induced metric given by
\begin{equation}
\tilde{G}^{ab} = \frac{\partial X^{M}}{\partial \sigma^a}
\frac{\partial X^{N}}{\partial \sigma^b} G_{MN}~.
\end{equation}
Assuming that $\tilde{G}_{ab}$ is non-degenerate, then one can
define the projection operator which projects vectors onto
tangent vectors to the submanifold $\Sigma$, i.e.,
\begin{equation}
\tilde{G}^{MN} = \frac{\partial X^{M}}{\partial \sigma^a}
\frac{\partial X^{N}}{\partial \sigma^b} \tilde{G}^{ab}~,
\end{equation}
which is known as the first fundamental tensor.
Furthermore one can define the projection operator onto the
vector space normal to $\Sigma$ by
\begin{equation}
\perp^{MN} = G^{MN} - \tilde{G}^{MN}~.
\end{equation}
It follows trivially that any normal to $\Sigma$, $n_M$ say,
satisfies $\perp^{MN} n_{N} = n^{M}$.  In particular this allows
one to show that $\perp^{MN}$ may be expanded as
\begin{equation}
\perp^{MN} = \sum_{i=p+1}^{D-p} n^{M}_i n^{N}_i~,
\end{equation}
for any orthonormal basis $n^{M}_i$ of normal vectors
to $\Sigma$.

In the so-called static gauge, ie., coordinates in which
$\Sigma$ can be parameterized
locally as $X^a = \sigma^a$, $X^i = X^i (\sigma^a)$, the 
induced metric takes the form (\ref{inducedG}) when the
metric is expanded about a flat background as 
$G_{MN} = \eta_{MN} + 2 \kappa H_{MN}$.  Inserting this
expansion into the definition of $\perp^{MN}$ one arrives
at
\begin{equation}
\perp^{MN} = D^{MN} - 4 \kappa D^{(M|P|} H_{PQ} N^{N)Q}
- 2 D^{(M|}_{\hspace{0.7em} P} 
\frac{\partial X^{P}}{\partial \sigma^a} N^{|N)a}~,
\end{equation}
for the expansion of $\perp^{MN}$ to linear order
in fluctuations.

The remaining geometrical quantity that we need to introduce
here is the second fundamental form.  Following \cite{Poly}
we define it\footnote{For an equivalent, but more covariant,
starting point, see \cite{Car}.}
through the decomposition
\begin{equation}
\frac{\partial X^M}{\partial \sigma^a} \nabla_M \Bigl(
\frac{\partial X^N}{\partial \sigma^b} \partial_N \Bigr)
= \tilde{\Gamma}^{c}_{ab} \Bigl( 
\frac{\partial X^{N}}{\partial \sigma^c} \partial_N \Bigr)
+ K^{i}_{ab} n_i~.
\end{equation}
This expression tells one how a tangent vector,
$\partial_b = (\partial X^N/\partial \sigma^b) \partial_N$,
changes under infinitesimal displacement along another
tangent, $\partial_a$, to $\Sigma$.  This change has 
components both tangent to the submanifold which defines
the connection coefficients, $\tilde{\Gamma}^{c}_{ab}$,
for the induced derivative operator on $\Sigma$, and
components normal
to the submanifold which defines the second fundamental
tensor $K^{i}_{ab}$.  Projecting this decomposition
onto a normal vector produces the component expression
\begin{equation}
K^{i}_{ab} = \Bigl(\frac{\partial^2 X^{N}}{\partial \sigma^a
\partial \sigma^b} + \frac{\partial X^{P}}{\partial \sigma^a}
\frac{\partial X^{Q}}{\partial \sigma^b} \Gamma^{N}_{PQ} \Bigr)
n^{i}_N~.
\end{equation}
Working in static gauge and expanding about the Minkowski metric
produces the expansion
\begin{equation}
K^{i}_{ab} = \kappa (\partial_a H_{b}^{\hspace{0.5em} i} +
\partial_b H_{a}^{\hspace{0.5em} i} - \partial^i H_{ab} )
+ \partial_a \partial_b X^i~.
\end{equation} 
Our form of the second fundamental form is related to that
in \cite{Car} by
\begin{equation}
K^{i}_{ab} = \frac{\partial X^M}{\partial \sigma^a}
\frac{\partial X^N}{\partial \sigma^b} K_{MN}^{\hspace{1.5em} P} n^{i}_P~.
\end{equation}

\end{document}